# Strongly enhanced charge-density-wave order in monolayer NbSe$_2$

Xiaoxiang Xi[1], Liang Zhao[2], Zefang Wang[1], Helmuth Berger[3], László Forró[3], Jie Shan[1,2*], and Kin Fai Mak[1*]

[1]Department of Physics, The Pennsylvania State University, University Park, Pennsylvania 16802-6300, USA
[2]Department of Physics, Case Western Reserve University, Cleveland, Ohio 44106-7079, USA
[3]Institute of Condensed Matter Physics, Ecole Polytechnique Fédérale de Lausanne, 1015 Lausanne, Switzerland
*E-mails: jus59@psu.edu; kzm11@psu.edu

**Two-dimensional (2D) atomic materials possess very different properties from their bulk counterparts. While changes in the single-particle electronic properties have been extensively investigated[1-3], modifications in the many-body collective phenomena in the exact 2D limit, where interaction effects are strongly enhanced, remain mysterious. Here we report a combined optical and electrical transport study on the many-body collective-order phase diagram of 2D NbSe$_2$. Both the charge density wave (CDW) and the superconducting phase have been observed down to the monolayer limit. While the superconducting transition temperature ($T_C$) decreases with lowering the layer thickness, the newly observed CDW transition temperature ($T_{CDW}$) increases drastically from 33 K in the bulk to 145 K in the monolayers. Such highly unusual enhancement of CDWs in atomically thin samples can be understood as a result of significantly enhanced electron-phonon interactions in 2D NbSe$_2$, which cause a crossover from the weak coupling to the strong coupling limit. This is supported by the large blueshift of the collective amplitude vibrations observed in our experiment.**

CDWs are periodic modulations of conduction electron densities and the associated lattice distortions in solids[4,5]. Understanding such spontaneous charge orders in 2D is essential to unravel the mechanism of unconventional superconductivity due to their intricate relationship in the phase diagram[6,7]. Dimensionality has profound effects on CDW instabilities as in many other phase transition phenomena[4]. On the one hand, reduced dimensionality strengthens Peierls instabilities (due to Fermi surface nesting) and electron-phonon interactions (due to reduced dielectric screening), favoring stronger CDWs[8]. On the other hand, stronger fluctuation effects from both finite temperatures and disorders tend to destroy long-range CDW coherence in low-dimensional systems[4]. Although long-range CDW coherence is well known in quasi-one-dimensional and quasi-two-dimensional systems[4], the interplay of these effects in the exact 2D limit remains unknown. Van der Waals' materials that can be separated into stable layers of atomic thickness provide an ideal platform for investigations of CDWs and their relation with superconductivity in the 2D limit[9,10].

Transition metal dichalcogenide niobium diselenide (NbSe$_2$) is one of the most studied van der Waals' materials that exhibit both CDW and superconductivity at low temperatures[11]. Monolayer NbSe$_2$ consists of an atomic layer of Nb sandwiched between two layers of Se in a trigonal prismatic structure (Fig. 1a). Bulk 2H-NbSe$_2$ is formed by stacking monolayers of NbSe$_2$ in the ABAB… sequence, where adjacent layers are rotated by 180° with respect to each other. A non-commensurate CDW order appears in bulk 2H-NbSe$_2$ below $T_{CDW}$ = 33.5 K as a second-order phase transition from the normal metal state[11]; and superconductivity emerges when the sample temperature is lowered further below $T_C$ = 7.2 K[12]. The CDW order in NbSe$_2$ is characterized by a three-component complex order parameter $\Delta_j = |\Delta_j|e^{i\phi_j}$ ($j$ = 1,2,3) with $|\Delta|$ and $\phi$ denoting the CDW gap and phase, respectively[13]. The precise origin of the CDW instability in 2H-NbSe$_2$ has been a topic of debate for many years[14-19]. Recent experimental and theoretical studies point to the momentum-dependent electron-phonon interactions[16-19], rather than the Fermi surface nesting[15] or saddle point singularities[14] as the major driving mechanism. While a new CDW order, distinct from that in the bulk 2H-NbSe$_2$, has been predicted by density functional theories in monolayer samples[8], experimental studies on the CDW instability in atomically thin NbSe$_2$ have not been reported.

Unlike superconductivity, the variation in the transport properties of a material undergoing a CDW transition (due to variation in the conduction electron density) is often weak[12]. The CDW transition, however, has strong optical signatures. A significant softening of the acoustic phonon mode at the CDW wavevectors occurs as temperature approaches $T_{CDW}$ from above, which then freezes for $T < T_{CDW}$[4,18]. Meanwhile, collective excitations of the CDW gap (amplitude mode) and phase (phase mode) emerge with the former being Raman active[4,20]. Here we have employed the terahertz-Raman spectroscopy (down to 8 cm$^{-1}$ Raman shift) to determine the CDW transition temperature $T_{CDW}$ in few-layer NbSe$_2$ by measuring the temperature dependence of both the soft mode and the amplitude mode. We have also performed complementary electrical measurements to obtain the corresponding superconducting transition temperature $T_C$ (See Method).

Figure 1b shows the optical image of a typical 2D NbSe$_2$ sample/device used in this study. Representative Raman spectra for bulk, bilayer and monolayer NbSe$_2$ at *room temperature* are shown in Fig. 1d for both the parallel (red lines) and perpendicular (blue lines) polarization configurations. Similar Raman spectra and polarization selection rules are observed for samples of different thickness. The prominent features observed below 300 cm$^{-1}$ include $E_{2g}$ modes at ~ 240, 180 and 20-30 cm$^{-1}$, and an $A_{1g}$ mode at ~ 220 cm$^{-1}$. The low-frequency mode $\omega_S$ = 20 - 30 cm$^{-1}$, which is strongly dependent on the layer number $N$ and absent in monolayers (Fig. 1e), originates from interlayer shearing. With decreasing $N$, the shear mode frequency $\omega_S$ decreases rapidly due to the reduced effective interlayer spring constant. It can be quantitatively described as $\omega_S = \omega_{S,Bulk}\cos(\frac{\pi}{2N})$ with a bulk shear mode frequency $\omega_{S,Bulk}$ = 29.4 cm$^{-1}$ (solid line, Fig. 1f). Such a thickness dependence of the shear mode frequency has been demonstrated for other van der Waals' materials[21]. It has been used to accurately determine the layer number $N$ of few-layer NbSe$_2$ in this study. In contrast, the three other Raman features are weakly dependent on $N$. The two high-energy modes (~ 220 and 240 cm$^{-1}$) are associated with in-plane phonons[20]. The mode around 180 cm$^{-1}$, as we discuss in more details below, is assigned as a soft mode which involves a second-order scattering process of two acoustic phonons of frequency $\omega_0$ at wavevector ~ $\frac{2}{3}\Gamma M$[20].

Figure 2 summarizes the temperature dependence of the Raman spectra for bulk, bilayer and monolayer NbSe$_2$ ranging from 6 - 180 K. The perpendicular polarization configuration is chosen to reduce the background from elastic scattering. The behavior of the bulk sample agrees very well with that reported in the literature[20,22]. Namely, the soft mode redshifts with decreasing temperature; it reaches ~ 115 cm$^{-1}$ and freezes for $T < T_{CDW}$ = 33.5 K. Below $T_{CDW}$ two new modes appear in the Raman spectra. The first is an amplitude mode (~ 35 cm$^{-1}$)[20,22], a collective excitation of the CDW fluctuations whose frequency ($\omega_A$) and intensity ($I_A$) both drop rapidly to zero when $T_{CDW}$ is approached from below (hence a second-order phase transition). The second new mode is a weak feature around 190 cm$^{-1}$ whose origin is unknown. In atomically thin NbSe$_2$, both the amplitude mode and the high-frequency weak feature, exclusive to the CDW order, are observed. While the frequency of the weak feature stays nearly independent of sample thickness $N$, the amplitude mode frequency $\omega_A$ (its zero-temperature value) increases monotonically from ~ 35 cm$^{-1}$ in the bulk to ~ 70 cm$^{-1}$ in monolayers (Fig. 2d-f). What's more striking is that both features survive up to much higher temperatures, indicating higher CDW transition temperatures. For more details on the Raman measurements refer to Supplementary Materials Section 1.

Several approaches have been employed to extract the CDW phase transition temperature $T_{CDW}$ of samples of different thickness and have yielded consistent results (Supplementary Materials Section 4). In Fig. 3a we show the determination of $T_{CDW}$ from the intensity of the amplitude mode $I_A$. For a given sample we first normalize its Raman spectra $S_T(\omega)$ at temperature $T$ by a reference spectrum $S_0(\omega)$ at $T_0 > T_{CDW}$. This allows us to eliminate effects that are temperature independent and thus unrelated with the CDW transitions. We then extract the integrated intensity of the amplitude mode as a function of temperature (Fig. 3a). The transition temperature $T_{CDW}$ is determined as the temperature at which $I_A$ crosses zero. The solid lines are fits to the mean field theory and serve as guides to the eye (Supplementary Materials Section 4.1). The method yields a bulk CDW transition temperature of $T_{CDW}$ = 36 ± 1 K which is consistent with published results. We summarize $T_{CDW}$ as a function of layer thickness $N$ in the phase diagram of Fig. 3b. With decreasing layer thickness, $T_{CDW}$ shifts towards higher temperatures and reaches 145 ± 3 K in the limit of monolayer thickness.

The significant enhancement of $T_{CDW}$ observed here in atomically thin NbSe$_2$ is highly unusual. The presence of collective excitations below $T_{CDW}$ indicates that the lateral phase coherence length (also referred to as the correlation length) is longer than the optical probe wavelength. $T_{CDW}$ extracted from our experiment is thus the temperature above which the quasi-long-range CDW order is destroyed. Contrary to our experiment, lowering the dimensionality of a system generally tends to suppress the long-range order and lower $T_{CDW}$ by the fluctuation effects since the available phase space for the phase parameter $\phi$ to adapt to imperfections (arisen from finite temperature and impurities) is much reduced[4]. Therefore, other effects such as the Fermi surface structure and the electron-phonon coupling strength in atomically thin samples must be considered to understand the measured phase diagram.

To this end, we perform a careful analysis of the soft mode frequency $\omega_0$ and the amplitude mode frequency $\omega_A$ in samples of different thickness (Supplementary Materials Section 2 and 3). The soft mode frequency at high temperature (300 K) shows a negligible dependence on $N$. This suggests that likely the same soft phonon mode is involved in driving the CDW order in atomically thin NbSe$_2$ as in the bulk. On the other hand, the amplitude mode frequency $\omega_A$, determined from the A-channel Raman spectra (obtained by subtracting the perpendicular from the parallel polarization configuration) (Fig. 4a), shows a significant blueshift with decreasing $N$, in accord with $T_{CDW}$. As we discuss below, this result, in combination with the Fermi surface similarity for samples of different thickness[8], suggests that the increasing electron-phonon coupling strength is likely the major driving mechanism for stronger CDWs in atomically thin NbSe$_2$.

In Fig. 4b we illustrate $T_{CDW}$ as a function of the normalized amplitude mode frequency $\lambda = \frac{\omega_A^2}{\omega_0^2}$, where $\omega_0$ is the un-renormalized soft phonon mode frequency[4] and its room temperature value of 88 cm$^{-1}$ (half of the experimental soft mode frequency) has been used. In the mean field theory, the dimensionless parameter $\lambda$ is the electron-phonon coupling constant[4]. We compare our experimental data with the mean field prediction for $T_{CDW}$ in the weak coupling limit $T_{CDW} = T_0 e^{-\frac{1}{\lambda}}$ ($\lambda \ll 1$). Here $T_0 = 35,548$ K has been used to match the bulk $T_{CDW}$. The value corresponds to Fermi energy $\epsilon_F = \frac{k_B T_0}{2.28} = 1.34$ eV ($k_B$ is the Boltzman constant), which is comparable with the result of ab initio calculations[8]. The weak coupling model (solid line, Fig. 4b) reproduces the trend of $T_{CDW}$ for thick samples, but fails to describe the experimental data for $N < 5$ (or $\lambda > 0.2$) with $T_{CDW} > 100$ K. The result suggests a crossover from a weak coupling (long coherence length Peierls-like instability) to a strong coupling regime (short coherence length local-bonding CDW)[23]. The upper bound temperature that corresponds to the crossover can be estimated from the Debye temperature[5,23] of NbSe$_2$ ($T_D$ = 220 K[24]): $T_D/1.76 \sim 125$ K (dashed line, Fig. 4b) which is consistent with the above picture.

Finally, we discuss superconductivity in atomically thin NbSe$_2$. We determine $T_C$ of samples of varying thickness by measuring the temperature dependence of their longitudinal electrical resistance $R$. Bulk NbSe$_2$ behaves as a good metal with a residual resistance ratio exceeding 30 before reaching the superconducting transition at $T_C$ = 7.2 K (Fig. 1c). A careful examination of $R(T)$ also shows a broad kink around 30 K corresponding to the CDW transition[12]. Similar behavior is observed in atomically thin samples with progressive suppression of $T_C$ to about 3.1 K in monolayers. (See Supplementary Materials Section 5 for more details.) The general trend observed here is consistent with earlier measurements available down to 2-3 layers[25,26]. The CDW transition, however, is no longer visible likely due to the broadening of the CDW feature. With the $N$-dependence of both $T_{CDW}$ and $T_C$, we complete the $N - T$ phase diagram in Fig. 3b. The two transition temperatures anti-correlate. Similar relation between $T_{CDW}$ and $T_C$ has also been observed in high-pressure studies of bulk 2H-NbSe$_2$[27]. It is attempting to assign the two opposite thickness dependences of the transition temperature to competing CDW and superconductivity orders. However, other effects such as reduced Josephson interlayer coupling[28] or/and enhanced electron-electron repulsion in atomically thin samples[29] can be evoked to explain the observed thickness dependence of $T_C$ in atomically thin NbSe$_2$. Further, while our experiment indicates the presence of strongly enhanced CDW order in atomically thin NbSe$_2$ driven by a crossover from weak to strong electron-phonon coupling, the use of the mean field analysis only provides an approximate picture for the problem. A more rigorous theoretical approach is needed. Meanwhile, measurements of the momentum dependence of the CDW gap are warranted for a complete understanding of the nature of the CDW order and its relationship to superconductivity in 2D NbSe$_2$.

**Methods**

**Sample preparation:** High-quality 2H-NbSe$_2$ single crystals were grown from Nb metal wires of 99.95% purity and Se pellets of 99.999% purity by iodine 99.8% vapor transport in a gradient of 730 °C – 700 °C in a sealed quartz tubes for 21 days. A very slight excess of Se was introduced (typically 0.2% of the charge) to ensure stoichiometry in the resulting crystals. Thin flakes were mechanically exfoliated from bulk single crystals on silicone elastomer polydimethylsiloxane (PDMS) stamps. Atomically thin samples were first identified by optical microscopy and then transferred to sapphire substrates for Raman measurements or silicon substrates (covered by a 280 nm layer of thermal oxide) with pre-patterned Au electrodes for electrical measurements (Fig. 1b). The sample thickness was determined according to their shear mode frequency by Raman spectroscopy as described in the main text. To minimize the environmental effects on the samples, we have limited their exposure to air to < 1-2 hours. For further protection, hexagonal boron nitride (hBN) thin films of 10's nm thickness have been introduced as a capping layer on NbSe$_2$ devices following the mechanical transfer method of Wang et al[30].

**Characterizations:** Terahertz-Raman spectroscopy was performed under normal incidence with a HeNe laser centered at 632.8 nm. The laser beam was focused to a diameter of about 1 $\mu$m onto samples by a 40 × objective. The reflected radiation was collected by the same objective and analyzed with a grating spectrometer equipped with a liquid nitrogen cooled charge coupled device (CCD). A combination of one reflective Bragg grating and two Bragg notch filters removes majority of the laser side bands and allows measurements of Raman shift down to ~ 8 cm$^{-1}$. The typical spectral resolution is 1.5 cm$^{-1}$. The absorbance in atomically thin NbSe$_2$ samples is about 0.026/layer at the excitation wavelength. To avoid significant laser heating of the samples, we have used excitation power of 0.12, 0.8, 1.2, and 1.2 mW (on the sample), respectively, for the bulk, trilayer, bilayer, and monolayer samples. The better thermal coupling to the substrates through interfacial heat diffusion in thinner samples allows us to use higher excitation laser powers. The heating effect is estimated to be < 0.5 K under the excitation conditions for all samples. An integration time of 20 minutes was used to obtain each spectrum of Fig. 2 and 4. The temperature dependence of the Raman spectra ranging from 6 - 180 K was measured in a Montana Instruments Cryostation. Resistivity measurements were carried out in a Physical Property Measurement System (PMMS) down to 2.1 K. Longitudinal electrical resistance was acquired using a four-point geometry.

**Figures and Figure Captions**

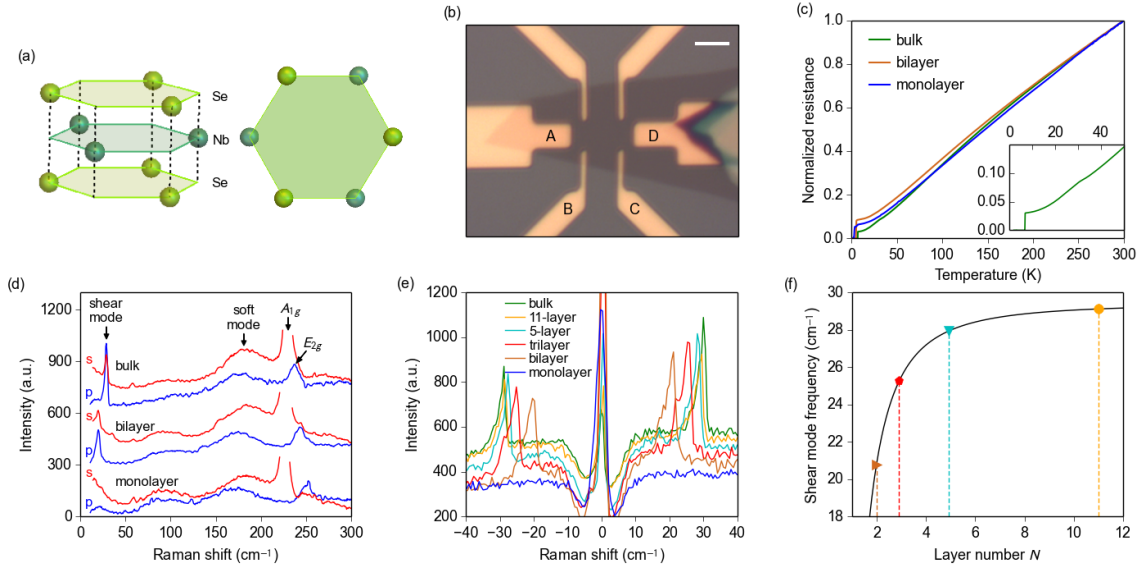

**Figure 1. Characterizations of atomically thin NbSe₂ samples.** (a) Trigonal prismatic structure of monolayer NbSe$_2$ (left) and the corresponding honeycomb lattice formed by the niobium and selenium sublattices (right). (b) Optical image of a bilayer NbSe$_2$ device. Current was passed through the sample via gold electrodes A and D; voltage drop was measured across B and C. The scale bar corresponds to 5 μm. (c) Temperature dependence of the resistance of bulk, bilayer and monolayer NbSe$_2$, normalized to their values at 300 K. The inset shows the bulk data below 50 K. (d) Raman spectra of bulk, bilayer, and monolayer NbSe$_2$ at room temperature for the parallel (s, red lines) and perpendicular (p, blue lines) polarization configurations. The $A_{1g}$ phonon peak is suppressed for the latter. (e) Raman spectra of bulk and atomically thin NbSe$_2$ at 170 K for the perpendicular polarization configuration show the Stokes and anti-Stokes lines of the shear mode for layer thickness $N > 1$. (f) Layer thickness $N$ dependence of the shear mode frequency (symbols). The solid line corresponds to $\omega_{S,Bulk} \cos\left(\frac{\pi}{2N}\right)$ with $\omega_{S,Bulk} = 29.4$ cm$^{-1}$. The spectra in (d) and (e) are vertically displaced for clarity.

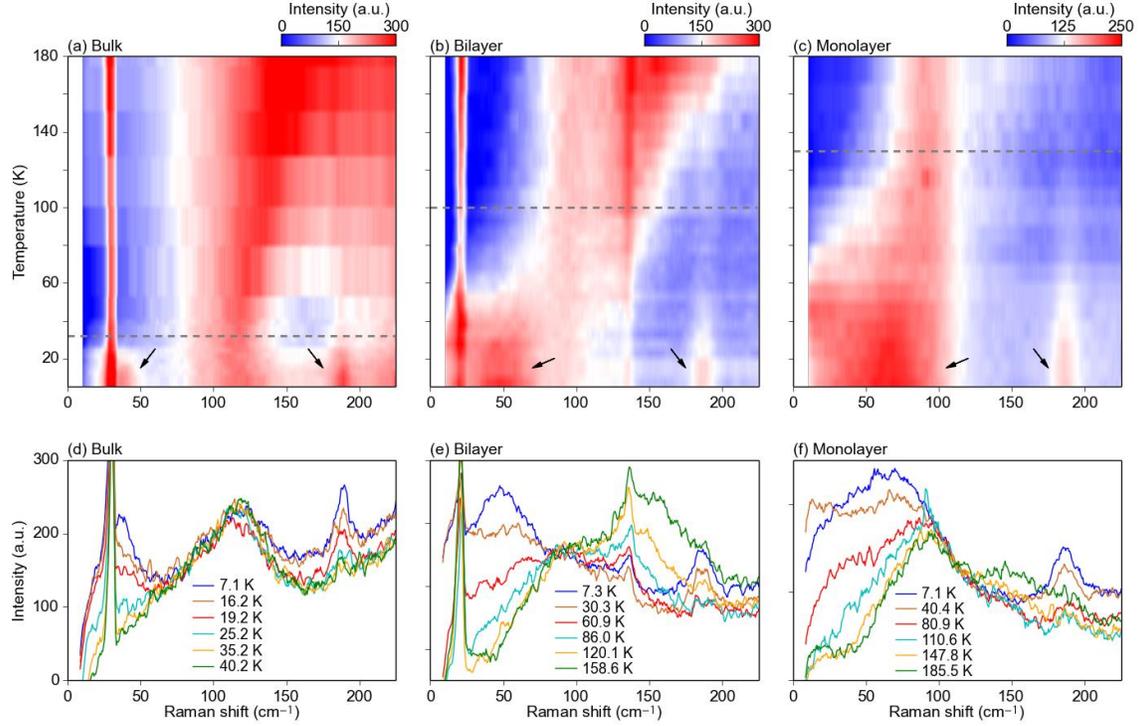

**Figure 2. Temperature dependence of Raman spectra of bulk and 2D NbSe$_2$.** (a) – (c) are temperature maps of Raman scattering intensity of bulk, bilayer, and monolayer NbSe$_2$, respectively, for the perpendicular polarization configuration. The two arrows in each map indicate the amplitude mode (the low frequency feature) and the weak high frequency feature. The dashed lines approximately delineate $T_{CDW}$. (d) - (f) are Raman spectra of each sample at selected temperatures. A constant background, which is dependent on temperature, has been subtracted from each spectrum.

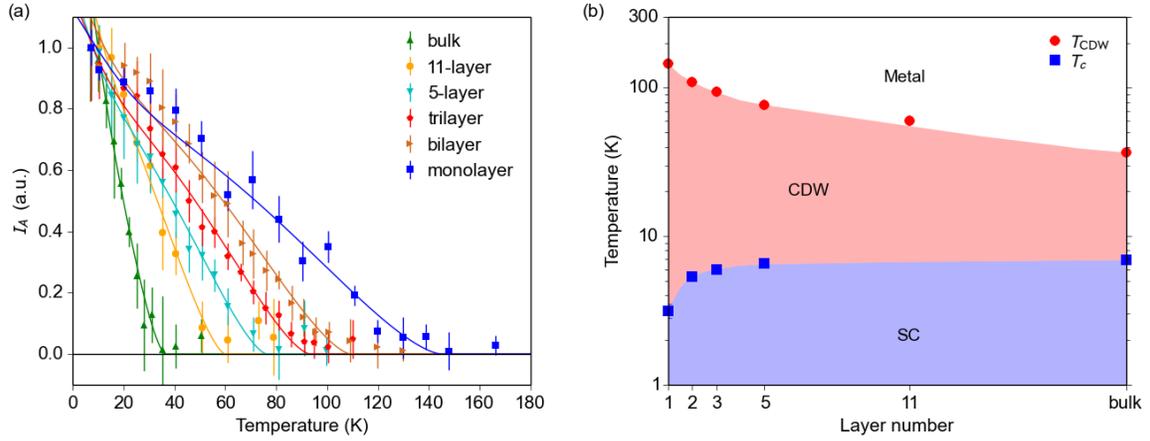

**Figure 3. Many-body phase diagram of 2D NbSe$_2$.** (a) Temperature dependence of the amplitude mode intensity $I_A$ for NbSe$_2$ samples of differing thickness. The solid lines are fits to the mean field theory. (b) $N - T$ phase diagram for NbSe$_2$. Symbols are the CDW and superconducting transition temperatures determined in this work with uncertainties about the size of the symbols. The boundaries between the phases are guides to the eye.

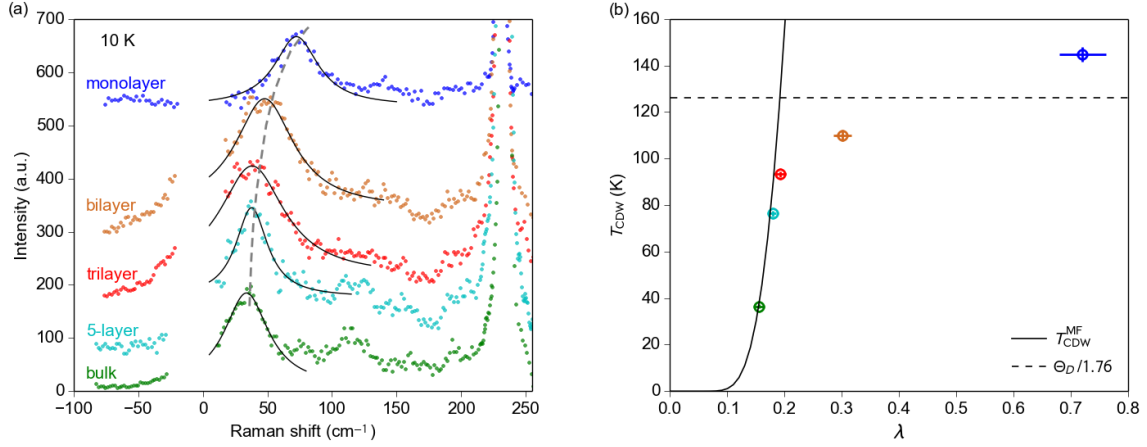

**Figure 4. Thickness dependence of the amplitude mode frequency and the electron-phonon coupling in NbSe₂.** (a) A-channel Raman spectra of NbSe$_2$ of varying thickness at 10 K (dotted lines). Spectra are vertically displaced for clarity. The solid lines are fits to a Lorentzian function. The dashed line illustrates the blueshift of the amplitude mode frequency $\omega_A$ with decreasing layer thickness. (b) $T_{CDW}$ as a function of normalized amplitude mode frequency $\lambda = \omega_A^2/\omega_0^2$ as defined in the text. The horizontal error bars are estimated from the uncertainties in the fit for $\omega_A$. The solid line is the weak-coupling mean-field prediction for the CDW transition temperature $T_{CDW}^{MF}$. The dashed line is an estimate based on the Debye temperature $\theta_D$ for a crossover from the weak to the strong coupling regime.

**References**

1. Geim, A. K. & Novoselov, K. S. The rise of graphene. *Nat Mater* **6**, 183-191 (2007).
2. Castro Neto, A. H., Guinea, F., Peres, N. M. R., Novoselov, K. S. & Geim, A. K. The electronic properties of graphene. *Reviews of Modern Physics* **81**, 109-162 (2009).
3. Wang, Q. H., Kalantar-Zadeh, K., Kis, A., Coleman, J. N. & Strano, M. S. Electronics and optoelectronics of two-dimensional transition metal dichalcogenides. *Nat Nanotechnol* **7**, 699-712 (2012).
4. Gruner, G. *Density Waves In Solids*. (Westview Press, 2009).
5. Rossnagel, K. On the origin of charge-density waves in select layered transition-metal dichalcogenides. *J Phys-Condens Mat* **23,** 213001 (2011).
6. Chang, J. *et al.* Direct observation of competition between superconductivity and charge density wave order in $YBa_2Cu_3O_{6.67}$. *Nat Phys* **8**, 871-876 (2012).
7. Joe, Y. I. *et al.* Emergence of charge density wave domain walls above the superconducting dome in $1T-TiSe_2$. *Nat Phys* **10**, 421-425 (2014).
8. Calandra, M., Mazin, I. I. & Mauri, F. Effect of dimensionality on the charge-density wave in few-layer $NbSe_2$. *Physical Review B* **80**, 241108 (2009).
9. Novoselov, K. S. *et al.* Two-dimensional atomic crystals. *P Natl Acad Sci USA* **102**, 10451-10453 (2005).
10. Yu, Y. *et al.* Gate-tunable phase transitions in thin flakes of $1T-TaS_2$. *Nat Nano* **advance online publication** (2015).
11. Wilson, J. A., Di Salvo, F. J. & Mahajan, S. Charge-density waves and superlattices in the metallic layered transition metal dichalcogenides (Reprinted from Advances in Physics, vol 32, pg 882, 1974). *Adv Phys* **50**, 1171-1248 (2001).
12. Soto, F. *et al.* Electric and magnetic characterization of $NbSe_2$ single crystals: Anisotropic superconducting fluctuations above TC. *Physica C: Superconductivity* **460–462, Part 2**, 789-790 (2007).
13. McMillan, W. Landau theory of charge-density waves in transition-metal dichalcogenides. *Physical Review B* **12**, 1187-1196 (1975).
14. Rice, T. M. & Scott, G. K. New Mechanism for a Charge-Density-Wave Instability. *Physical Review Letters* **35**, 120-123 (1975).
15. Straub, T. *et al.* Charge-Density-Wave Mechanism in $2H-NbSe_2$: Photoemission Results. *Physical Review Letters* **82**, 4504-4507 (1999).
16. Valla, T. *et al.* Quasiparticle Spectra, Charge-Density Waves, Superconductivity, and Electron-Phonon Coupling in $2H-NbSe_2$. *Physical Review Letters* **92**, 086401 (2004).
17. Kiss, T. *et al.* Charge-order-maximized momentum-dependent superconductivity. *Nat Phys* **3**, 720-725 (2007).
18. Weber, F. *et al.* Extended Phonon Collapse and the Origin of the Charge-Density Wave in $2H-NbSe_2$. *Physical Review Letters* **107**, 107403 (2011).
19. Arguello, C. J. *et al.* Quasiparticle Interference, Quasiparticle Interactions, and the Origin of the Charge Density Wave in $2H-NbSe_2$. *Physical Review Letters* **114**, 037001 (2015).

20	Tsang, J. C., Smith, J. E. & Shafer, M. W. Raman Spectroscopy of Soft Modes at the Charge-Density-Wave Phase Transition in 2H-NbSe2. *Physical Review Letters* **37**, 1407-1410 (1976).
21	Tan, P. H. *et al.* The shear mode of multilayer graphene. *Nat Mater* **11**, 294-300 (2012).
22	Sooryakumar, R. & Klein, M. V. Raman Scattering by Superconducting-Gap Excitations and Their Coupling to Charge-Density Waves. *Physical Review Letters* **45**, 660-662 (1980).
23	Mcmillan, W. L. Microscopic Model of Charge-Density Waves in 2h-Tase2. *Physical Review B* **16**, 643-650 (1977).
24	Harper, J. M. E., Geballe, T. H. & Disalvo, F. J. Thermal-Properties of Layered Transition-Metal Dichalcogenides at Charge-Density-Wave Transitions. *Physical Review B* **15**, 2943-2951 (1977).
25	Frindt, R. F. Superconductivity in Ultrathin NbSe2 Layers. *Physical Review Letters* **28**, 299-301 (1972).
26	Staley, N. E. *et al.* Electric field effect on superconductivity in atomically thin flakes of NbSe2. *Physical Review B* **80** (2009).
27	Chu, C. W., Diatschenko, V., Huang, C. Y. & DiSalvo, F. J. Pressure effect on the charge-density-wave formation in 2H-NbSe2 and correlation between structural instabilities and superconductivity in unstable solids. *Physical Review B* **15**, 1340-1342 (1977).
28	Schneider, T., Gedik, Z. & Ciraci, S. From Low to High-Temperature Superconductivity - a Dimensional Crossover Phenomenon - a Finite Size Effect. *Z Phys B Con Mat* **83**, 313-321 (1991).
29	McMillan, W. L. Transition Temperature of Strong-Coupled Superconductors. *Physical Review* **167**, 331-344 (1968).
30	Wang, L. *et al.* One-Dimensional Electrical Contact to a Two-Dimensional Material. *Science* **342**, 614-617 (2013).

# Supplementary Materials
# Strongly enhanced charge-density-wave order in monolayer NbSe$_2$

Xiaoxiang Xi[1], Liang Zhao[2], Zefang Wang[1], Helmuth Berger[3], László Forró[3], Jie Shan[1,2*], and Kin Fai Mak[1*]

[1]Department of Physics, The Pennsylvania State University, University Park, Pennsylvania 16802-6300, USA
[2]Department of Physics, Case Western Reserve University, Cleveland, Ohio 44106-7079, USA
[3]Institute of Condensed Matter Physics, Ecole Polytechnique Fédérale de Lausanne, 1015 Lausanne, Switzerland
*E-mails: jus59@psu.edu; kzm11@psu.edu

## 1. Optical characterization of atomically thin NbSe$_2$

### 1.1 Optical contrast of atomically thin NbSe$_2$ on sapphire substrates

For Raman spectroscopy, NbSe$_2$ samples on transparent sapphire substrates were used to reduce scattering background. Fig. S1 is an optical micrograph of samples of varying thickness on a sapphire substrate. The thickness of samples was first determined by their optical contrast and then confirmed independently by their shear mode frequency as described in the main text and atomic force microscopy (AFM).

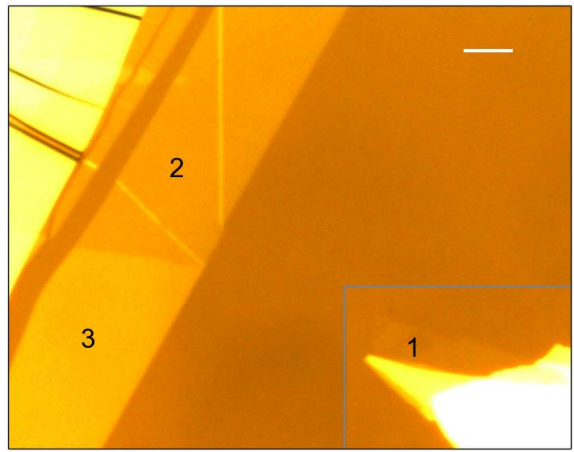

Figure S1. Optical image of ultrathin NbSe$_2$ crystals on a sapphire substrate. Monolayer, bilayer, and trilayer regions are identified. The scale bar corresponds to 5 μm.

### 1.2 Optical absorption spectrum of atomically thin NbSe$_2$

Absorption spectrum of atomically thin NbSe$_2$ samples in the energy range of 1.45 - 2.90 eV was measured using the method described in Ref. 3. It relies on the reflectance contrast of samples on transparent substrates. Fig. S2 is the absorption spectrum of monolayer NbSe$_2$ at 7 K. The absorbance is about 0.026 at the excitation wavelength (632.8 nm) used for Raman spectroscopy in this work. The value scales linearly with layer thickness and has a negligible temperature dependence.

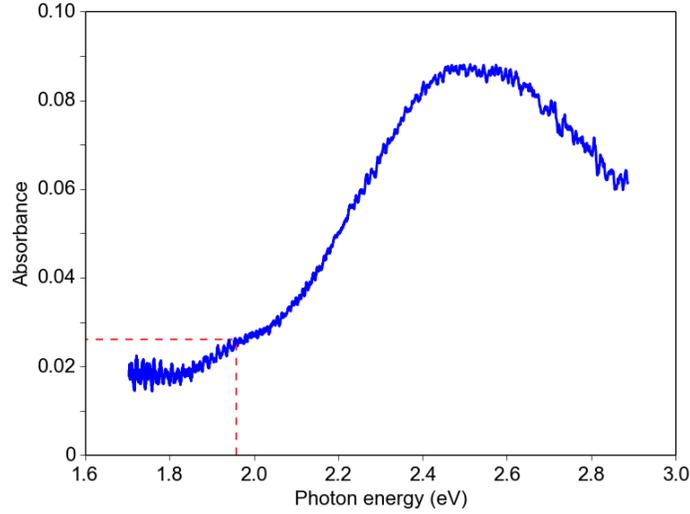

Figure S2. Absorption spectrum of monolayer $NbSe_2$ at 7 K ranging from 1.45 - 2.90 eV. The dashed lines indicate the excitation laser energy and the corresponding absorbance for Raman spectroscopy in this work.

### 1.3 Polarization dependence of Raman spectra

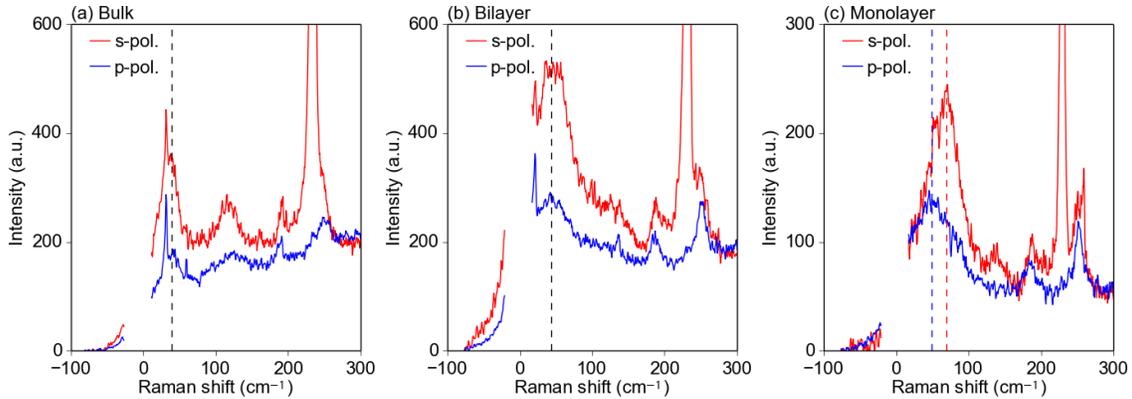

Figure S3. Comparison of the Raman spectra for the parallel (s) and perpendicular (p) polarization configurations for (a) bulk, (b) bilayer, and (c) monolayer $NbSe_2$ at 10 K. The data have been corrected for the difference in the collection efficiency of the two polarizations. Small contributions from the sapphire substrate have also been subtracted. Data near zero Raman shift are not shown due to the presence of the strong laser line. The dashed lines indicate the peak of the amplitude mode.

We measured Raman spectra of samples of different thickness for both the parallel (s) and perpendicular (p) polarization configurations. For a quantitative comparison, the data collection efficiency for the two polarizations was calibrated over the measured spectral range using a white light source. Sapphire substrates give rise to small scattering, which was measured independently and subtracted from all Raman spectra.

Representative data after these corrections are shown in Fig. S3 for samples in the charge-density-wave (CDW) phase. From the bulk down to the bilayers, the low energy peak below 100 cm$^{-1}$, which we assign as the amplitude mode, is situated at roughly the same frequency for both s- and p-polarization. For the monolayer sample, the peak for the s-polarization is at ~70 cm$^{-1}$ and at ~ 20 cm$^{-1}$ lower for the p-polarization. The weak feature near 190 cm$^{-1}$ associated with the CDWs is peaked at the same frequency for both polarizations for all samples. The difference spectra of the two polarizations (the A channel response) are presented in Fig. 4(a) of the main text for samples of different thickness and in Fig. S9(h) for a monolayer sample at different temperatures.

### 1.4 Sample dependence of Raman spectra

The main features in the Raman scattering spectra, particularly the amplitude mode, were reproducible for all samples of the same thickness studied in this work. Figure S4 shows the A channel Raman spectra of bulk, bilayer, and monolayer NbSe$_2$ at 10 K for 2 samples each. Samples prepared from the same batch on the same substrate are labeled with the same batch number.

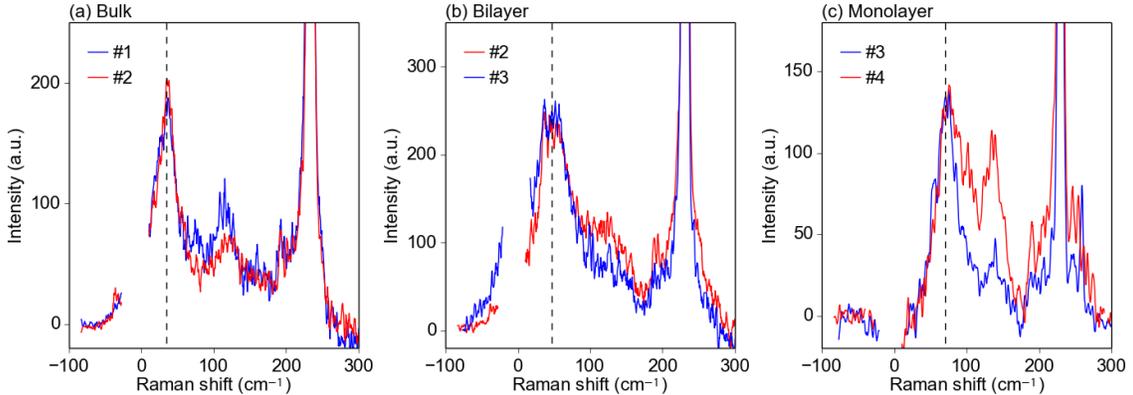

Figure S4. Sample variations of (A channel) Raman spectra of bulk (a), bilayer (b), and monolayer NbSe$_2$ (c) at 10 K. The dashed lines indicate approximately the amplitude mode peak frequency. Samples prepared on the same substrate are labeled by the same batch number.

### 2. Temperature dependence of the soft mode

It has been established that in bulk NbSe$_2$ the acoustic phonon mode near $\frac{2}{3}\Gamma M$ is associated with the CDW transition. Upon cooling, it softens significantly until reaching $T_{CDW}$, below which it is frozen. The mode around 180 cm$^{-1}$ observed in the Raman spectrum of bulk samples at room temperature [Figure S5(a)] corresponds to a second-order scattering process of the soft phonon. It is clearly discernable from 295 K down to 7 K and follows the established temperature dependence. For atomically thin samples, this mode is also present at room temperature [main text, Fig. 1(d)]. Upon cooling it softens, shown in Figure S5(b) and (c) for 5-layer and monolayer samples, respectively. The mode becomes weak in atomically thin samples at low temperatures. We were not able to trace its temperature dependence across the entire CDW phase transition. We compare the available temperature dependence of the soft mode frequency in samples of different thickness in Figure S5(d). The data indicate that samples of different thickness have the same soft mode frequency at room temperature, which suggests that the CDW phase transition for NbSe$_2$ samples of different thickness (down to monolayer) likely involves the same soft phonon mode, at variance with a theoretical prediction[4].

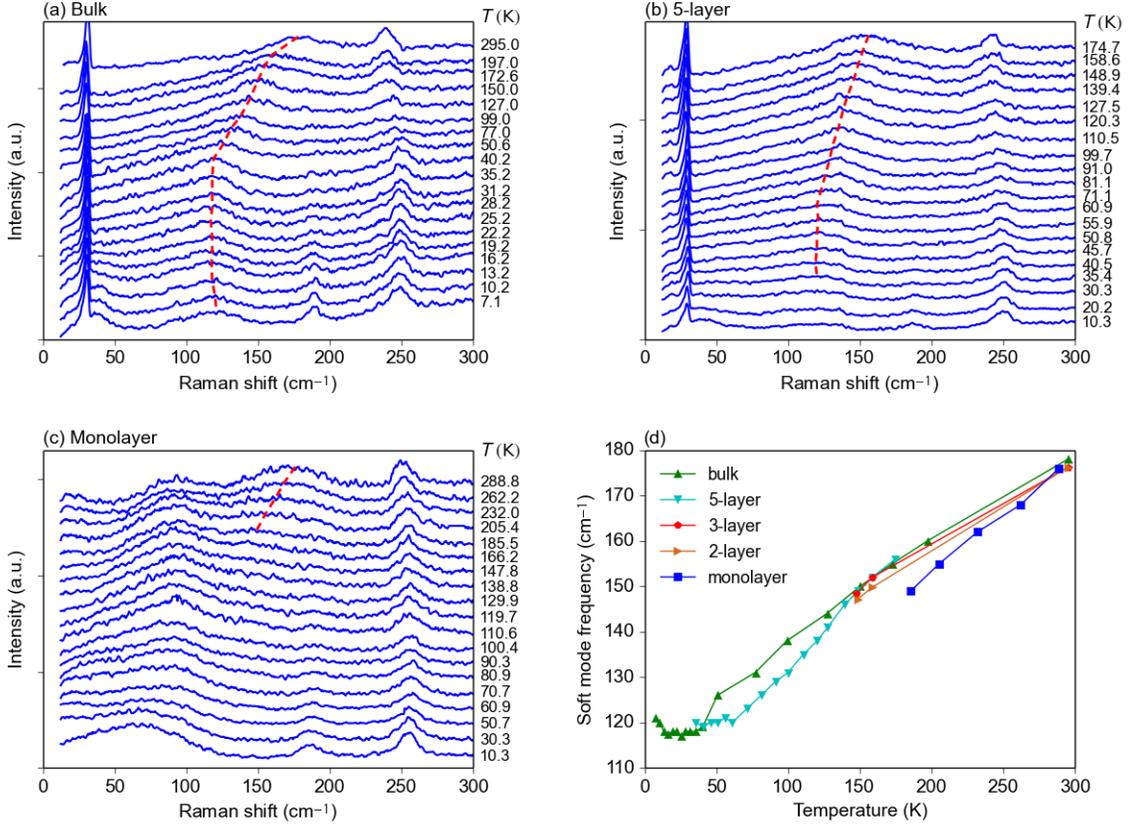

Figure S5. Raman spectra of (a) bulk, (b) 5-layer, and (c) monolayer NbSe$_2$ for the perpendicular polarization configuration. Spectra at different temperatures are displaced vertically for clarity. The dashed lines are guides to the eye for the soft mode peak frequency. (d) Temperature dependence of the soft mode frequency.

## 3. Determination of the amplitude mode frequency

The amplitude mode frequency $\omega_A$ was determined from the A channel response $\Delta S(\omega)$, i.e. the difference calculated from the raw spectra of the s- and p-polarizations (Fig. S6). This procedure minimizes the contributions in the Raman spectra of the shear mode and extrinsic effects such as defects. We used a Lorentzian function with a constant background B

$$\Delta S(\omega) = \frac{A}{\pi} \cdot \frac{\frac{1}{2}\gamma}{(\omega - \omega_A)^2 + (\frac{1}{2}\gamma)^2} + B. \qquad (1)$$

to describe the A channel response. The fitting results are shown as black lines in Fig. S6. The values of the center frequency $\omega_A$ and its uncertainty $\Delta\omega_A$ obtained from the fit are listed in Table S1. The table also includes the values for the normalized amplitude mode frequency $\lambda = \omega_A^2/\omega_0^2$ and its uncertainty $\Delta\lambda$ calculated from $\omega_A$ and $\omega_0 = 88$ cm$^{-1}$.

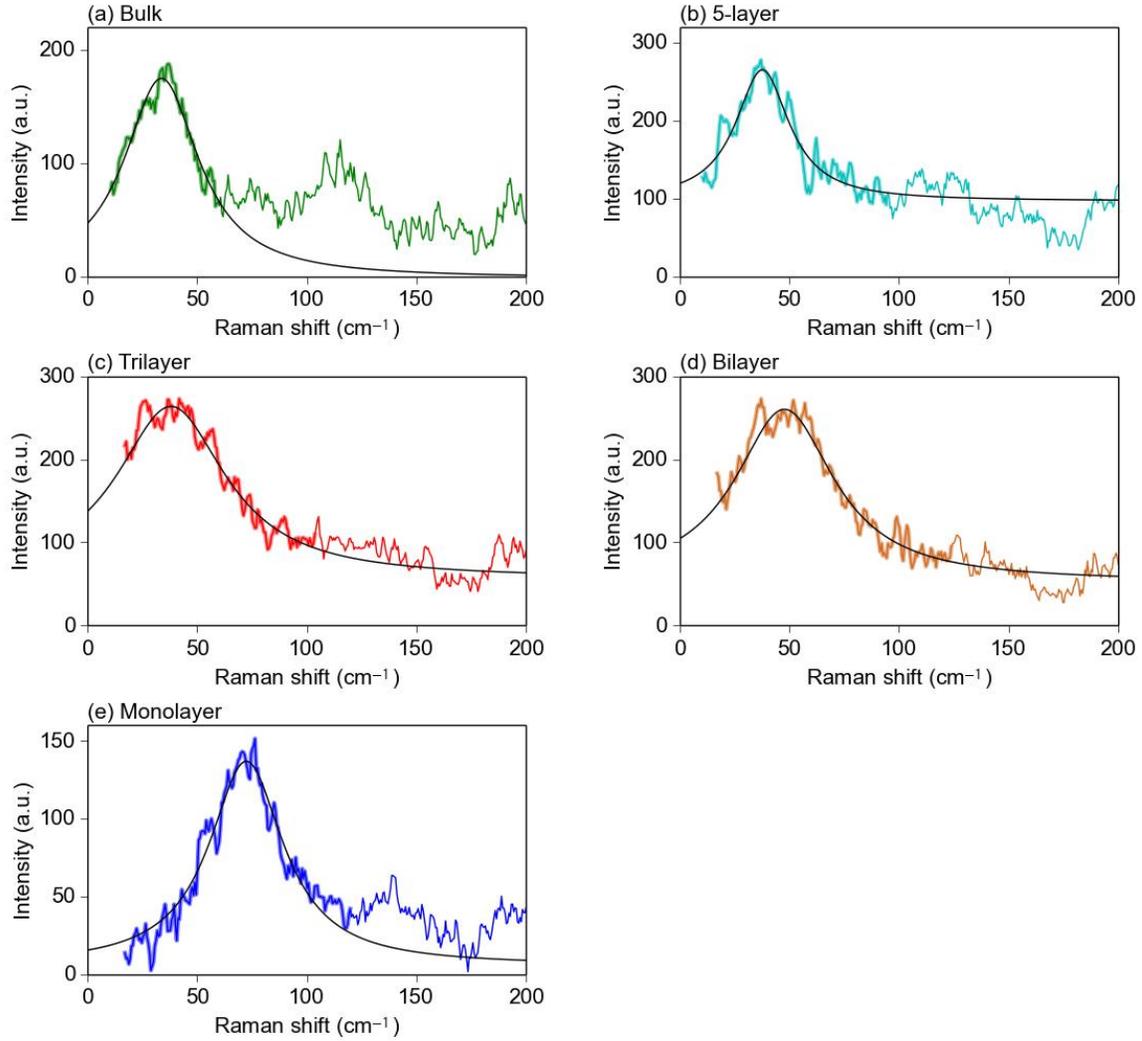

Figure S6. Analysis of the amplitude mode at 10 K for a bulk (a), 5-layer (b), trilayer (c), bilayer (d), and monolayer (e) sample. The colored lines are the difference spectra calculated from the raw spectra of the s- and p-polarization. The black lines are fits to Equation 1. The spectral range shown in thick lines was used in the fit.

Table S1. Layer number dependence of the amplitude mode frequency.

| Layer number | $\omega_A$ (cm$^{-1}$) | $\Delta\omega_A$ (cm$^{-1}$) | $\lambda = \omega_A^2/\omega_0^2$ | $\Delta\lambda$ |
|---|---|---|---|---|
| bulk | 34.7 | 1.0 | 0.155 | 0.009 |
| 5 | 37.4 | 0.6 | 0.181 | 0.006 |
| 3 | 38.6 | 1.0 | 0.192 | 0.010 |
| 2 | 48.3 | 1.3 | 0.301 | 0.016 |
| 1 | 74.7 | 2.1 | 0.721 | 0.041 |

## 4. Determination of $T_{CDW}$

In this section, we describe the analysis of $T_{CDW}$ based on the amplitude mode and the weak feature around 190 cm$^{-1}$ that are associated with CDW. For the monolayers, we also compare Raman scattering data measured using different polarization configurations.

### 4.1 Intensity of the amplitude mode $I_A$

In order to analyze the temperature dependence of the amplitude mode quantitatively and to extract the phase transition temperature $T_{CDW}$, we normalize all the Raman spectra $S_T(\omega)$ to a high temperature Raman spectrum $S_0(\omega)$ when the sample is in the normal phase ($T_0 > T_{CDW}$). This allows us to eliminate effects that are unrelated with the CDW transitions. The spectrum $S_0(\omega)$ can be chosen unambiguously because the weak feature ~ 190 cm$^{-1}$ completely disappears in the high temperature normal phase. The normalized spectra $S_T(\omega)/S_0(\omega)$ at different temperatures are shown in Fig. S7(a-c) for the bulk, bilayer and monolayer samples, respectively. The normalized intensity of the amplitude mode $I_A$ as a function of temperature for samples of different thickness (Fig. 3a of the main text) is obtained by integrating $S_T(\omega)/S_0(\omega)$ over the amplitude mode spectral range (indicated by dashed lines) at each temperature.

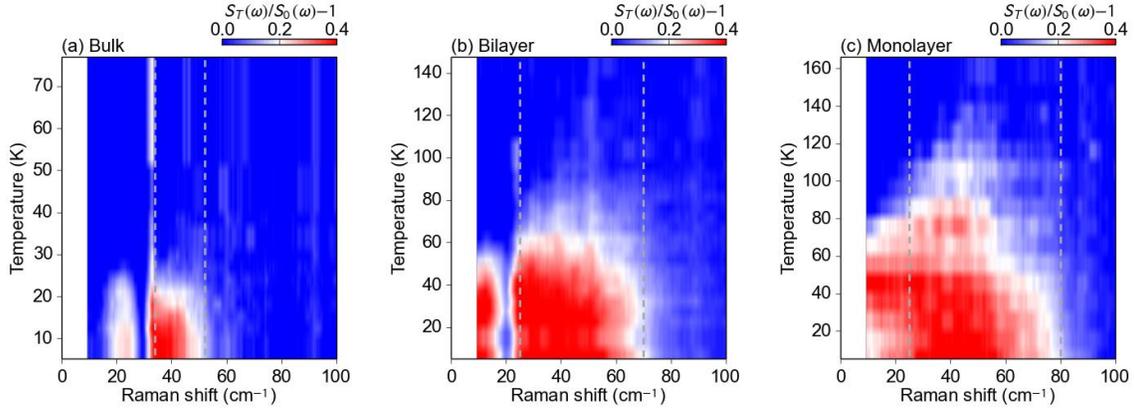

Figure S7. Temperature dependent Raman spectra $S_T(\omega)$ of (a) bulk, (b) bilayer, and (c) monolayer NbSe$_2$ normalized to their normal phase spectra $S_0(\omega)$ at 99.0 K, 147.5 K, and 185.5 K, respectively. The regions between the two dashed lines are used for calculation of $I_A$.

The temperature dependence of $I_A$ is given by[5]

$$I_A \propto \frac{|\Delta|^{2q}[n(\omega_A)+1]}{\omega_A}. \tag{2}$$

Here $\Delta$ is the amplitude of the order parameter; $q = 2$ for NbSe$_2$ is an exponent that is material dependent; $\omega_A$ is the amplitude mode frequency; and $n(\omega_A)$ is the Bose-Einstein distribution. Using the mean field results of $|\Delta| \propto (1 - \frac{T}{T_{CDW}})^{\frac{1}{2}}\Theta(T_{CDW} - T)$ and $\omega_A \propto (1 - \frac{T}{T_{CDW}})^{\frac{1}{4}}\Theta(T_{CDW} - T)$ with step function $\Theta(x)$, we fit the data in Fig. 3a of the main text to Equation 2, obtaining an estimate of $T_{CDW}$. The fits are shown as solid lines in Fig. 3a of the main text.

### 4.2 Analysis of the weak mode at ~ 190 cm$^{-1}$

The weak feature at ~ 190 cm$^{-1}$ is another signature of the CDW phase. Its strength decreases with increasing temperature and diminishes approximately at the phase transition, allowing an estimate of $T_{CDW}$. Its peak profile can be roughly decomposed as a Lorentzian function sitting on a linear background [shown as dashed lines in Fig. S8(a – c)]. The resultant fits are shown as solid blue lines in Fig. S8(a–c). $T_{CDW}$ were estimated by fitting the temperature dependence of the Lorentzian peak area $I_W$ to Equation 1 [solid lines in Fig. S8(d)]. $T_{CDW}$ obtained by analyzing the weak feature at ~ 190 cm$^{-1}$ is plotted in Fig. S8(e) and compared with the result obtained based on the amplitude mode. These results are consistent, illustrating the strongly enhanced $T_{CDW}$ in atomically thin NbSe$_2$ samples. The discrepancy between the two analyses may be attributed to several reasons, among which the most notable is the large background near the weak feature ~ 190 cm$^{-1}$.

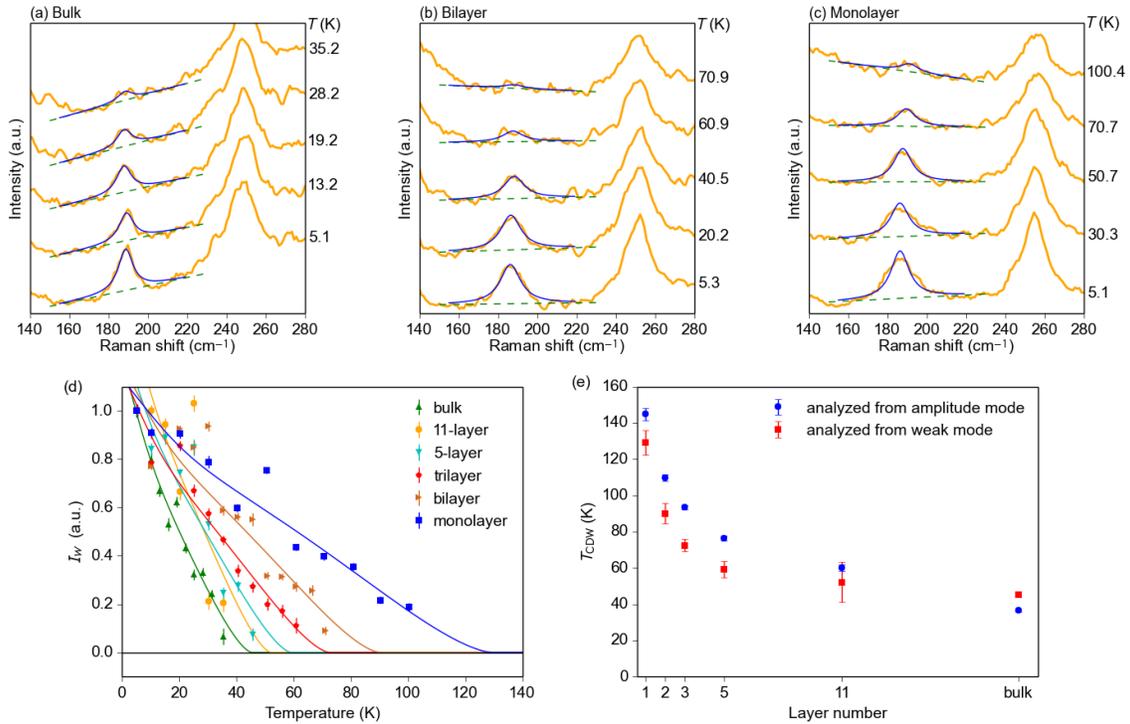

Figure S8. Analysis of the weak feature at ~ 190 cm$^{-1}$ for (a) bulk, (b) bilayer, and (c) monolayer NbSe$_2$. The yellow solid lines are experimental data, the dashed green lines are linear fits to the background, and the blue solid lines are fits combining the background and a Lorentzian. (d) Temperature dependence of the integrated area of the weak feature around ~ 190 cm$^{-1}$ $I_W$ for NbSe$_2$ samples of various thickness. The error bars are the standard deviations of $I_W$ from the fits. The solid lines are fits to Equation 2. (e) Layer number dependence of $T_{CDW}$ obtained from the analysis of the amplitude mode (circles) and from the weak feature at ~ 190 cm$^{-1}$ (squares).

### 4.3 Data of monolayer NbSe$_2$ for different polarizations

The temperature maps of Raman scattering intensity of Fig. S9(b – d) compare the Raman spectra of NbSe$_2$ monolayer sample #1 for the p- and s-polarizations, as well as the A channel response. Line cuts at selected temperatures are shown in Fig. S8(f – h). For comparison, the p-polarization data for monolayer sample #2 presented in Fig. 2 of the main text are included in Fig. S8(a) and (e) for comparison. All data show that the low-energy peak (below 100 cm$^{-1}$) persists to above 100 K. We used the same procedure as described in Section 4.1 to obtain the peak area for data in Fig. S8(a – c). For the difference spectra of Fig. S8(d), we integrated the peak area directly without normalizing the spectra to a high-temperature spectrum. The extracted $T_{CDW}$ are compared in Fig. S10. The deviations of $T_{CDW}$ obtained from different analyses are within ± 4%.

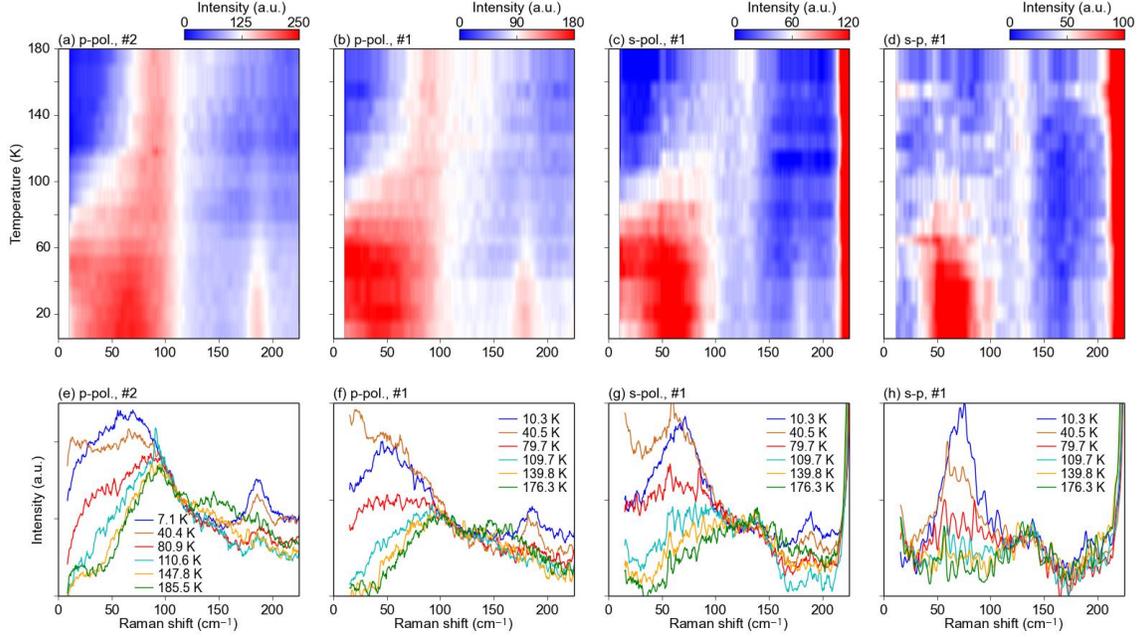

Figure S9. Temperature maps of Raman scattering intensity for monolayer NbSe$_2$ (a) Sample #2, p-polarization, (b) Sample #1, p-polarization, (c) Sample #1, s-polarization, and (d) Sample #1, difference between the s- and p-polarization. (e – h) are the corresponding Raman spectra at selected temperatures.

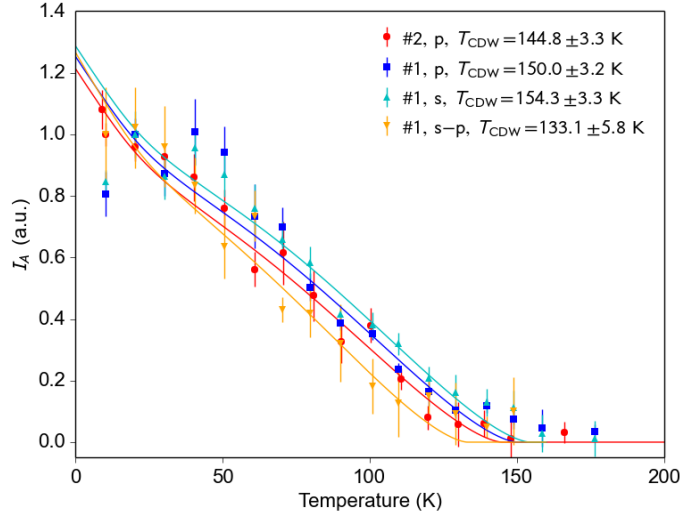

Figure S10. Temperature dependence of the amplitude mode area $I_A$ for monolayer NbSe$_2$ analyzed from the data in Figure S9. The solid lines are fits to Equation 2, yielding the $T_{CDW}$ shown in the legend.

## 5. Electrical characterization of atomically thin NbSe$_2$

To determine the superconducting transition temperature $T_C$ of NbSe$_2$ samples of various thickness, we fabricated devices using the method described in the Method Section of the main text. Four-point geometry was adopted for an accurate measurement of the sample resistance (Fig. S11). Multiple devices were prepared and measured. All yielded consistent results for samples of the same thickness. The superconducting transition temperature $T_C$ shown in Fig. 3b of the main text is determined as the temperature at which the resistance drops to 50% of the normal-state value.

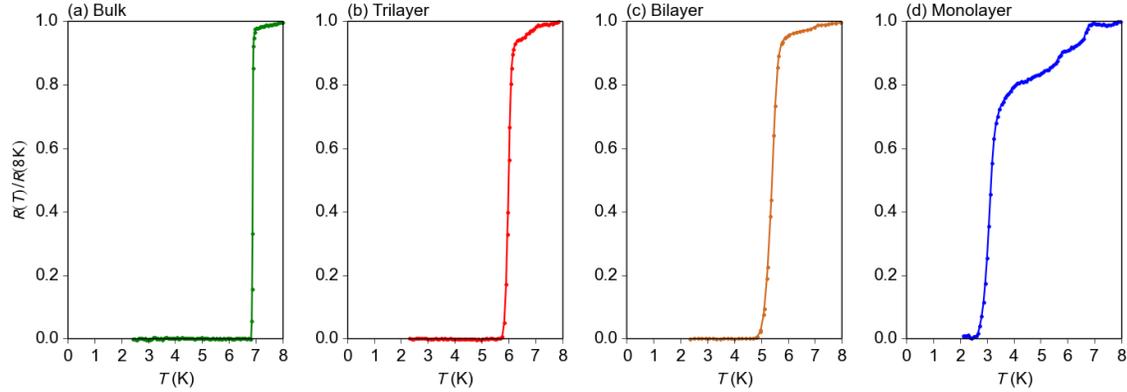

Figure S11. Temperature dependence of the resistance for (a) bulk, (b) trilayer, (c) bilayer and (d) monolayer NbSe$_2$, normalized to their values above the superconducting transition.

## References

1. Wang, L. *et al.* One-Dimensional Electrical Contact to a Two-Dimensional Material. *Science* **342**, 614-617 (2013).
2. Mak, K. F., Lui, C. H. & Heinz, T. F. Measurement of the thermal conductance of the graphene/SiO2 interface. *Applied Physics Letters* **97**, 221904 (2010).
3. Mak, K. *et al.* Measurement of the Optical Conductivity of Graphene. *Physical Review Letters* **101**, 196405 (2008).
4. Calandra, M., Mazin, I. I. & Mauri, F. Effect of dimensionality on the charge-density wave in few-layer NbSe2. *Physical Review B* **80**, 241108 (2009).
5. Tsang, J. C., Smith, J. E. & Shafer, M. W. Raman Spectroscopy of Soft Modes at the Charge-Density-Wave Phase Transition in 2H-NbSe2. *Physical Review Letters* **37**, 1407-1410 (1976).